# Electron and Hole Photoemission Detection for Band Offset Determination of Tunnel Field-Effect Transistor Heterojunctions

Wei Li,[1,2,*] Qin Zhang,[1,*] R. Bijesh,[3] Oleg A. Kirillov,[1] Yiran Liang,[2] Igor Levin,[1] Lian-Mao Peng,[2] Curt A. Richter,[1] Xuelei Liang,[2,a)] S. Datta,[3] David J. Gundlach[1] and N. V. Nguyen[1,a)]

1. National Institute of Standards and Technology, Gaithersburg, MD 20899, USA.
2. Key Laboratory for the Physics and Chemistry of Nanodevices and Department of Electronics, Peking University, Beijing 100871, China.
3. Department of Electrical Engineering, The Pennsylvania State University, University Park, Pennsylvania 16802, USA.

*These authors contributed equally to this work.
Contacts: Nhan.Nguyen@nist.gov; liangxl@pku.edu.cn;

The electrical performance of a tunnel field-effect transistor depends critically on the band offset at their semiconductor heterojunction interface.  Historically, it has been difficult to experimentally determine how the electronic bands align at the heterojunction interface. We report here on experimental methods to ascertain a complete energy band alignment of a broken-gap tunnel field-effect transistor based on an InAs/GaSb hetero-junction.  By using graphene as an optically transparent electrode in a traditional internal photoemission measurement, both the electron and hole barrier heights at the InAs/GaSb interface can be quantified.   For a $Al_2O_3$/InAs/GaSb layer structure, the barrier height from the top of InAs and GaSb valence band to the bottom of $Al_2O_3$ conduction band is inferred from electron emission whereas hole emissions reveal the barrier height from the top of $Al_2O_3$ valence band to the bottom of InAs and GaSb conduction band.  Subsequently, the offset parameter at the broken gap InAs/GaSb interface is extracted and thus can be used to facilitate the development of predicted model of electron quantum tunneling efficiency and transistor performance.

Different transistor designs for beyond-CMOS technology have been proposed including the tunnel field-effect transistor (TFET),[1] impact-ionization MOS,[2] ferroelectric FET,[3] and electromechanical devices.[4] The prototypes of these devices have been shown to achieve sub-threshold swing less than the 60 mV/dec intrinsic limit of current CMOS. Among the candidate designs, the TFET is considered a technologically promising candidate because it offers a much improved on-off current ($I_{ON}/I_{OFF}$) ratio over a given gate voltage swing and low power consumption, when compared with other candidates for the same performance.[1] The principle behind such advancement is the adoption of quantum-mechanical band-to-band tunneling (BTBT) as the switching mechanism, instead of thermionic emission which governs conventional CMOS operation in the subthreshold regime. Because BTBT can achieve steeper sub-threshold slope, operation at lower supply voltages and, thus, less power dissipation can be realized with TFETs.

In a TFET, the on-off current switching is controlled by applying a gate bias that shifts the channel valence band above the source conduction band so that carriers can tunnel into empty states of the channel. The most critical challenge in TFET design is to achieve high $I_{ON}$ and low $I_{OFF}$, and while maintaining a sub-threshold slope of less than 60 mV/decade over the largest possible range in the subthreshold regime. By various new designs of the gate dielectric and channel configuration of all-silicon TFETs, e.g. SOI-TFET, researchers have demonstrated low $I_{OFF}$, but still encountered a low $I_{ON}$.[5,6] The low tunneling current ($I_{ON}$) for Si-based TFETs is due to the relatively large bandgap and high tunneling effective mass.[7] Modest improvements have been achieved when replacing Si with $Si_{1-x}Ge_x$ on insulating substrates.[8] Given these fundamental limitations, it is reasonable to expect that Si-based TFETs will continue to suffer from low $I_{ON}$.[7]

An obvious solution is to use lower band gap materials and low-effective-mass materials, and take advantage of band engineering to increase BTBT. In fact, high $I_{ON}$ at lower voltages was achieved on Ge,[9] InAs, and in graphene nano-ribbons,[10,11] as well as heterojunction systems such as SiGe/Si,[12] AlGaSb/InAs,[13,14] AlGaAs/InGaAs,[15] and InGaSb/InGaAs.[16] Among these different designs, group III–V heterojunctions are considered to be very promising since they offer small effective masses and their band gaps can be tailored for desired band-edge alignments. Experimental and theoretical studies indicate that the performance of group III-V staggered or broken gap TFETs can be significantly enhanced when compared with homojunctions.[7,17,18] Because $I_{ON}$ depends on transmission probability over the interband tunneling barrier, which is a function of band offsets, band bending and other physical parameters at the source and channel interface, it is vital to design a device with appropriate heterojunction band offsets. Thus, having an accurate evaluation or measurement of the band offsets is critical to selecting a priori suitable heterojunction materials that will produce the necessary interfacial energy band edge arrangement.

Both band discontinuity and built-in potential determine carrier injection or confinement in a heterojunction device. The built-in potential caused by the band bending and the Fermi level at the interface is usually probed by electrical measurement of a Schottky barrier.[19,20] The extent of the band bending is on the order of the Debye length, which can range from a few angstroms to several hundred nanometers depending on the bulk doping of the two semiconductors. In contrast, the band discontinuities are the offsets due to high electrostatic potential gradients on the length scale of a single atomic inter-planar spacing for abrupt interfaces or a gradual variation of the local electronic density of states on the scale of only a few atomic layers; and they are mostly independent of doping. Earliest determination of semiconductor heterojunction band offsets come from transport measurements which is essentially a space-averaging technique whereas the band discontinuity is confined locally at the interface. The reliability of the transport measurements may be questionable due to the effects of contact quality, unintentional doping, leakage currents, and the likely input of unrealistic physical parameters to the modeling processes. Photoemission spectroscopy, such as X-ray photoemission (XPS), is capable of probing the local states by means of inelastic mean free path of photo excited electrons, but the sampling depth is limited to only a few nanometers from the surface. It is arguable that XPS is a reliable offset determination technique but the measurement results can be affected by chemical shifts, band bending, strain, and other intrinsic limitations.[21] To circumvent these limitations, optical techniques such as absorption, luminescence, light scattering, and internal photoemission (IPE) spectroscopies are often used.[21] Of these methods, IPE, which is based on transport mechanism of photoexcited carriers emitted over the interface barrier,[22] has been shown to be a robust, accurate, and direct measurement that offers some advantages drawn from both other optical and transport techniques.

In recent years, IPE has been successful in determining barrier height at a solid/solid interface, in particular, semiconductor/insulator and metal/insulator interfaces.[22] For heterojunction of narrow band gap semiconductors, due to small band offsets, IPE usually requires high intensity light sources such as free-electron laser which is not easily accessible.[23] Despite this constraint, conventional IPE on a TFET structure has been shown to be successful in quantifying band alignments of InGaAs/InAs and InAs/$p^+$AlGaSb.[24, 25] In these instances, an IPE experimental procedure was specifically designed and tailored to enhance sensitivity of electron photoemission from each semiconductor component of the heterojunction over a large band gap insulator. In both examples, the measurement technique was possible only when the larger band gap semiconductor is on top of the other. In this article, we advance the method to a more elaborate approach which simultaneously resolves both valence and conduction band offsets at the heterojunction interface without the restriction of the band gap arrangement used in Zhang's reports.[24, 25] Specifically the offset of the valence bands is determined by electron photoemission whereas that of conduction bands is measured by hole photoemission. Since hole photoemission is difficult to detect, we use graphene as a unique transparent electrode to enhance the hole emission. The proof of concept of the latter process has been validated on a less complicated test structure of $SiO_2$/Si where holes photo-injected from Si to $SiO_2$ yield the band offset of their valence band by using graphene on insulator.[26] This measurement strategy is adopted in this investigation to provide a complete band alignment of a semiconductor heterojunction. Another advantage of being able to draw a complete band alignment at the interface is, as a by-product, to extract the band gap of either the insulator or the semiconductor if one of them is known in advance. This also allows a consistency check of the IPE band offset measurement by comparing the extracted electrical band gap with the optical band gap independently measured by other techniques such as spectroscopic ellipsometry. The particular structure in this study consists of heterojunction of InAs/GaSb whose interface energy band alignment was designed to be a type III or broken-gap.  It is projected that this type of interface may lead to high frequency operation at much lower voltages presenting the potential of this emerging device technology for a range of power constrained applications such as distributed sensor networks, implantable medical electronics and ultra-mobile computing applications.  Therefore, accurate determination of the energy band offsets of this type of TFET heterojunction is contingent and critical to the tunneling efficiency for the corresponding device that strongly depends on the broken-gap offset.  Finally, this novel measurement approach is expected to be readily applied to other staggered and broken-gap heterojunction of TFET's and other beyond-CMOS devices.

**RESULTS AND DISCUSSSION**

Fig. 1(a) displays a schematic of the TFET InAs/GaSb heterojunction used in the IPE measurement where a bias $V_g$ is applied across the structure and photocurrent (Ig) is measured as a function of photon energy (hv) of incident light.  The IPE quantum yield (Y) is defined as a ratio of photocurrent and incident photon flux.  The aim of the measurement is to obtain the barrier height at the buried InAs/GaSb interface.  Monolayer graphene is employed as an optically semitransparent electrode to detect both IPE hole and electron photo-injections. The thickness of InAs was carefully designed to control light absorption and penetration depth in the layer stack and still to maintain the same pseudomorphism at the interface.  The high-angle annular dark field image acquired by using scanning transmission electron microscopy (STEM) verifies the layer thickness and interface sharpness (Fig. 1b).

Shown in Figs. 2(a) and 2(c) are the cube roots of IPE quantum yield, $Y^{1/3}$, versus photon energy when applied $V_g > V_{FB}$ where $V_{FB}$ is the flatband voltage, for the heterojunction with a 29 nm and 10 nm InAs layer, respectively.  The flatband voltage is equivalent to an externally applied potential at which the photocurrent switches the polarity or when the internal electric field in the oxide layer becomes zero.  The electric field across the oxide is estimated by $(V_g-V_{FB})/(Al_2O_3$ thickness) assuming the voltage drops entirely inside the oxide.[24]  Following the classical Powell model,[27] $Y^{1/3}$ is a linear function of photon energy above and near the spectral threshold for semiconductor/insulator interface.  The electrons escaping over the oxide conduction band under light illumination are photo-excited in the InAs and/or GaSb layer.  Discerning the source of material where these electrons emerge from can be carried out by observing whether the photoemission quantum yield contains optical absorption features belonging specifically to that material.  For semiconductors, the most common and unique features in the visible and ultraviolet part of optical absorption spectrum are associated with the inter-band transition critical points (CPs).[28]  The CPs of InAs and GaSb are recognized from their dielectric functions shown in Fig. 2(b), which are measured by spectroscopic ellipsometry.[22, 29, 30]  CPs relevant to IPE data interpretation are those of $E_0'$, $E_0' + \Delta_0'$, and $E_2$ transitions indicated in Fig. 2(b) of the imaginary part ($\varepsilon_2$) of the dielectric functions.  The IPE yield from 29 nm InAs sample contains $E_0'$ ( ~ 4.4 eV) and $E_2$ (~ 4.6 eV) being direct transitions from the valence band to the conduction band at the $\Gamma$ and X point of the Brillouin zone, respectively.[31]  With a high optical absorption in the range of $E_0'$ and $E_2$ point it is expected that the quantum yield will be enhanced.  However, in the vicinity of $E_0'$, the yield increases faster with increasing photon energy and deviates from the preceding linear region; displaying a bump. At both of these two CPs, the quantum efficiency shows a different trend, whereas at $E_2$, the yield remains unchanged. These different trends can be explained by how the band structure at $E_0^{'}$ and $E_2$ of InAs lines up with the large band gap of $Al_2O_3$. Fig. 3 displays a schematic of the band structure of InAs that is so arranged in relation with the valence and conduction band of $Al_2O_3$. $E_0^{'}$ associates with the direct interband transition from the top of valence band at Γ point to the bottom of higher conduction band (indicated by vertical line $E_0^{'}$ in Fig. 3) indicating the photo-excited electrons in final state of higher energy above the conduction band edge of $Al_2O_3$, thereby, contributing to and enhancing the photo-electron yield. On the contrary, the final state (indicated by vertical line $E_2$ in Fig. 3) of photo-excited electrons at the X point lies below the conduction band edge of $Al_2O_3$ thus contributing no photoelectrons to the IPE yield. Furthermore, none of GaSb CPs appears in the photoemission spectrum since, for photon energies near the barrier height threshold, the incidence light is mostly absorbed in the 29 nm InAs layer and less than 5% of incident light can penetrate into GaSb for photon energy larger than 2.7 eV.[25] Therefore it can be safely concluded that the photocurrents originate mainly from the InAs layer. In contrast, Fig. 2(c) presents the IPE $Y^{1/3}$ for much thinner (10 nm) InAs layer sample which allows most light transmitted into the GaSb layer. It contains three CPs ($E_0^{'}$, $E_0^{'} + \Delta_0^{'}$, and $E_2$) corresponding to GaSb absorption feature, where $E_0^{'}$ and $E_0^{'} + \Delta_0^{'}$ correspond to direct gap transitions and the spin-orbit split at the Γ point in the Brillouin zone, respectively, and the $E_2$ feature is due to transitions along Σ or near the X point.[32, 33] Thus we deduce that the photocurrents are due to photoemission from the GaSb. From these observations, it is concluded that the lower thresholds in Fig. 2(a) are the barrier heights from the InAs valence band maximum to the $Al_2O_3$ conduction band minimum and the higher thresholds in Fig. 2(c) are the barrier heights from the GaSb valence band maximum to the $Al_2O_3$ conduction band minimum.

The barrier heights for electron photoemissions ($V_g > V_{FB}$), extracted from Figs. 2(a) and 2(c) are observed to be field dependent due to the image force lowering effect.[19] The lowering appears to be a greater effect for the InAs layer adjacent to the oxide and lesser for the farther GaSb layer. The flatband or *zero*-field barrier height ($\Phi_0$) can be determined by the linear relationship of Schottky plot of barrier height vs. square-root of electric field as shown in Figure 4. As a result, $\Phi_0$ from the InAs and GaSb valence band maximum to the $Al_2O_3$ conduction band minimum is determined to be 3.45 eV and 2.92 eV with a 0.05 eV uncertainty. The band offset at InAs and GaSb interface can be deduced from their conduction band offsets with respect to the valence band of $Al_2O_3$. They can be conclusively determined by measuring the corresponding hole barrier heights.

In a traditional IPE measurement, the hole photocurrent (if present) from semiconductor is negligible due to its much lower quantum yield compared to the electron photocurrent from the semi-transparent gate (usually thin metal).[34] However, Yan *et al.*[35] first reported that by using graphene as transparent electrode one can greatly enhance the detection sensitivity to hole photoemission. The main advantages of using graphene as transparent electrode to facilitate hole photocurrent measurements are described in the following. In conventional IPE measurements, the electron photoemission from a thin layer of metal which is used as a semi-transparent electrode normally overwhelms the hole emission from the semiconductor substrate. Replacing the metal with a monolayer graphene, the broad range light transmittance as high as 97.7% [36] allows most of the incident photon flux to reach the emitter thus minimizing electron injection from the graphene electrode,[37] and increasing the external quantum efficiency of the hole emission. In addition, the resistivity of pristine graphene has been estimated to be as low as $10^{-6}$ Ω·cm, which is lower than silver electrode, and the sheet resistance of monolayer graphene can be 30Ω/□ at room temperature, which is comparable to highly conducting transparent electrode such as indium tin oxide.[38] The high electrical conductivity of graphene makes the collection of the emitted carriers more efficient and decreases carrier recombination. As a result, shown in Fig. 5(a) are photon current yield of the hole emissions for thick InAs layer sample. Since this layer absorbs most of the incident light, the observed photo-excited hole emission comes from the InAs layer thus the threshold corresponds to the barrier height from the InAs conduction band to the $Al_2O_3$ valence band. Unlike electron photoemission thresholds, hole photoemission thresholds appear to be field-independent. Further theoretical investigation should be taken to explain this observation. A similar independence on electric field has been observed in other material systems.[26] Consequently, from Fig. 5(a) the field-independent band offset from InAs conduction band minimum to the $Al_2O_3$ valence band maximum is found to be 3.20 eV. In the case of thin InAs layer sample, mainly hole emission from GaSb layer is observed, and the barrier height from GaSb conduction band minimum to the $Al_2O_3$ valence band maximum is 4.10 eV as shown in Fig. 5(b). The IPE yield spectrum from thick InAs layer sample features a signature of InAs as shown by the absorption peak $E_0^{'}$ in Fig. 2(b). At the photon energy of this critical point, the hole emission is enhanced and can be associated with the direct $E_0^{'}$ optical excitation of InAs. On the other hand, the plateau seen in the quantum yield near 4.3 eV (see Fig. 5(b)) corresponding to the $E_2$ feature, a transition along Σ or near the X point of GaSb, may indicate a lesser contribution of the excited holes to photo emission yield because their final state may lie below the valence band edge of $Al_2O_3$.

From the *zero*-field barriers determined in the above band diagram of thick and thin InAs layer, the InAs/GaSb broken-gap hetero-junction can be schematically established as shown in Fig. 6(a) and 6(b). It is interesting to verify the consistency of the IPE barrier height results by comparing the band gap of $Al_2O_3$ of 6.29 eV calculated from the band alignment with the band gap of 6.30 eV independently determined from optical absorption measurement on the same $Al_2O_3$.[39] Finally, the broken gap of ∼ -0.18 eV between the conduction band edge of InAs and the valence band edge of GaSb is extracted from the band diagram in Fig. 6.

**Conclusions**

In summary, we demonstrate the utility of IPE measurements to quantitatively characterize both the electron and hole barrier heights in the heterojunction of a TFET. Taking advantage of the high transmissivity and conductivity of monolayer graphene and using it as a transparent electrode for IPE measurements, we are able to detect holes photo-injected over an interface barrier. By sequentially measuring the electron and hole photoemission currents we are able to determine the energetic barrier heights at the heterojunction interface, and derive the complete and quantitative electronic band alignment. The knowledge infrastructure established here provides critical physical input parameters to facilitate the design and advancement of heterojunction TFETs. The methodology reported here to construct the band alignment of InAs and GaSb broken-gap heterojunction are broadly applicable to other herojuction materials systems and device technologies, e.g. solar cells.

**Methods**

**Device Fabrication**. The fabrication method for the InAs/GaSb broken gap semiconductor heterojunction reported here has been previously described in detail elsewhere.[16] We specifically select a thick (29 nm) and thin (10 nm) InAs layer in order to detect photoemission separately from each layer of the heterojunction. A similar sample with a much thinner InAs was used to check for consistency in IPE data interpretation. A 15 nm thick $Al_2O_3$ gate dielectric was deposited by atomic layer deposition at 110°C by using trimethylaluminum and water as precursors. Next, chemical vapor deposition (CVD) grown monolayer graphene was transferred onto the sample by using a "modified RCA clean and transfer method".[40-43] $100 \times 200$ μm$^2$ rectangular graphene regions were patterned by using photolithography and oxygen plasma etching. Finally, a 200 nm thick pad of aluminum was deposited by e-beam evaporation to form a mechanically robust and electrically reliable contact for the IPE measurement.

**Internal Photoemission**. Details of the IPE measurement setup are as described by Nguyen et al.[44] The IPE photocurrents were measured as a function of photon energy from 1.5 eV to 5.0 eV with applied gate bias ($V_g$) varied from -1.5 V to 1.5 V (or from -1.0 V to 1.0 V) in steps of 0.1 V. The IPE yield was calculated by the ratio of the photocurrent to the incident light flux. The oxide electric field is calculated from the offset of the applied bias $V_g$ from the built-in flat-band voltage ($V_{FB}$). $V_{FB}$ is derived from the applied bias at which the photocurrent near the emission threshold switches direction from positive to negative.[30] For both samples, $V_{FB}$ is determined to be 0.6 V with respect to the grounded substrate.

**Acknowledgement**:
This work was supported by the Ministry of Science and Technology of China (Grant No. 2011CB921904) and the Ministry of education of China (Grant No. 113003A). W. L. was partly supported by the National Institute of Standards and Technology. The authors would like to acknowledge NIST NanoFab's support for device fabrications.

# Figure captions

**Figure 1**. (a) Schematic of the IPE measurement of the graphene/Al$_2$O$_3$/InAs/GaSb structure used in this investigation. The top contact for IPE probe is a thick aluminum. The heterojunction of interest is InAs/GaSb; (b) the high-angle annular dark-field STEM image of this heterostructure confirming the layer thickness and interface sharpness.

**Figure 2**. (a) and (c) are the cube root of the photocurrent yield ($Y^{1/3}$) as a function of photon energy at different gate biases applied between GaSb substrate and aluminum contact for two graphene/ Al$_2$O$_3$/InAs/GaSb structures, one with 29 nm thick InAs layer and the other 10 nm thick, (b) is the imaginary part $\langle\varepsilon_2\rangle$ of the pseudo-dielectric function of InAs (orange) and GaSb (magenta) measured by spectroscopic ellipsometry.

**Figure 3** The energy band structure of InAs (image reprinted with permission from Ref. 31) is arranged to align with the band edge of Al$_2$O$_3$ to illustrate the yield modulations. At $E_0^{'}$ critical point, the photo-excited electrons in final state of energy higher than the conduction band edge of Al$_2$O$_3$ add more to the photo-current and thus enhance the photo-electron yield. The final state at $E_2$ lies below the conduction band edge of Al$_2$O$_3$ contributes no photoelectrons to the IPE yield.

**Figure 4**. Schottky plot of the barrier height as a function of square root of oxide electric field. Dash line is a linear fit to determine the zero-field barrier height ($\Phi_0$) at the oxide flat band condition. $\Phi_0$ = 3.45 eV ± 0.05 eV and 2.92 eV ± 0.05 eV is the barrier height from the InAs and GaSb valence band maximum to the Al$_2$O$_3$.

**Figure 5**. The cube root of the photocurrent yield due to hole emission from (a) InAs and (b) GaSb as a function of photon energy at different gate biases. A linear fit near the yield threshold results in a band offset of 3.20 eV from InAs conduction band minimum to the Al$_2$O$_3$ valence band maximum for the thick InAs. For the thin InAs layer sample, the onset of hole emission at 4.10 eV ascertains the barrier height from GaSb conduction band minimum to the Al$_2$O$_3$ valence band maximum.

**Figure 6**. The band alignment (not to scale) of broken-gap InAs and GaSb heterojunction at the oxide zero field: (a) thick InAs and (b) thin InAs.

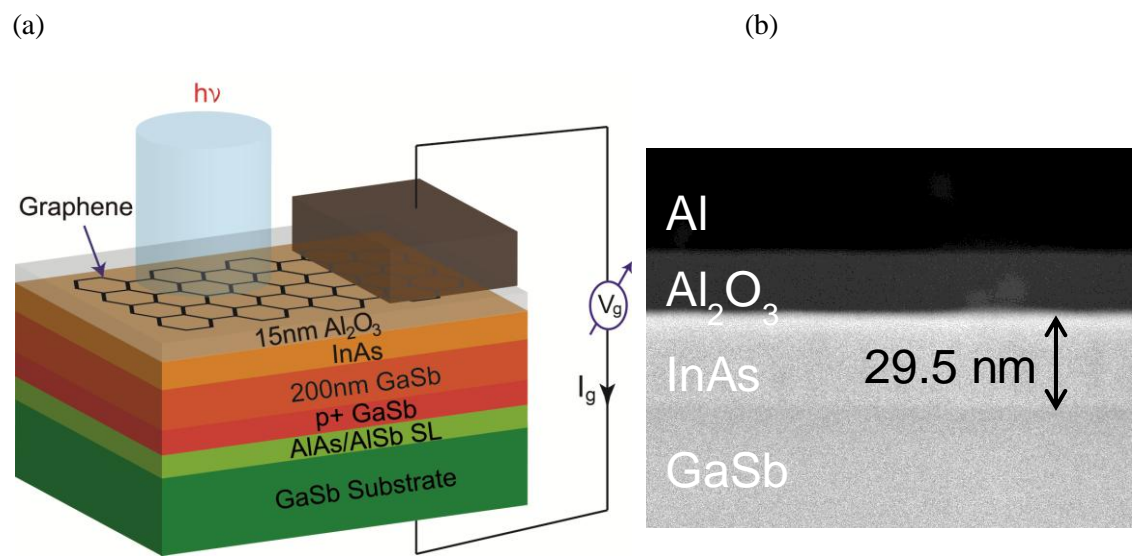

Figure 1

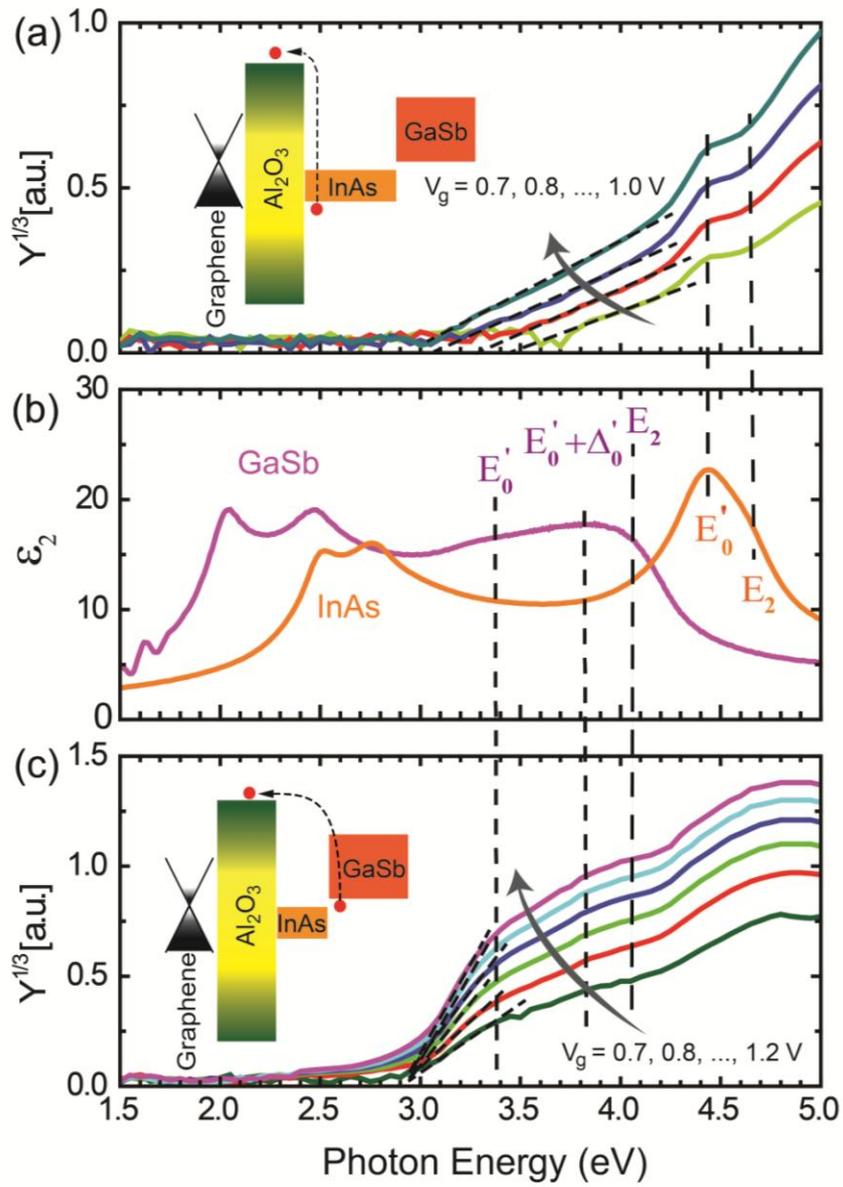

Figure 2

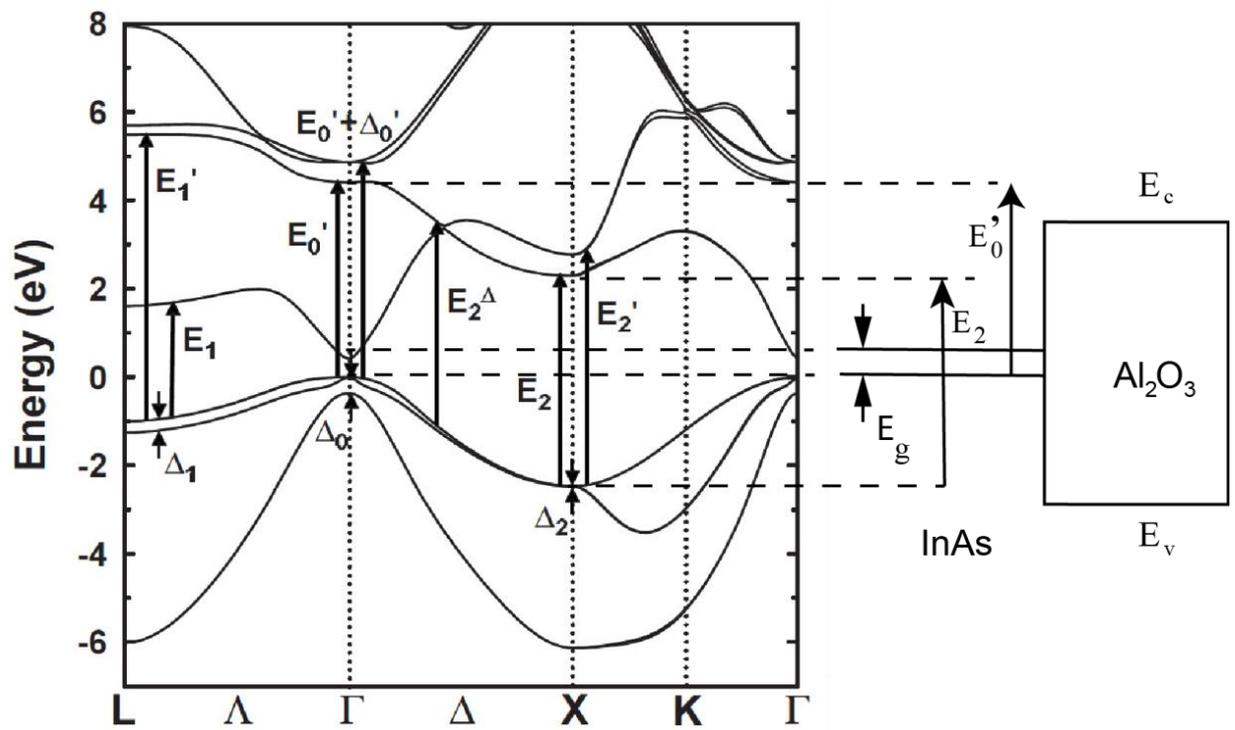

Figure 3

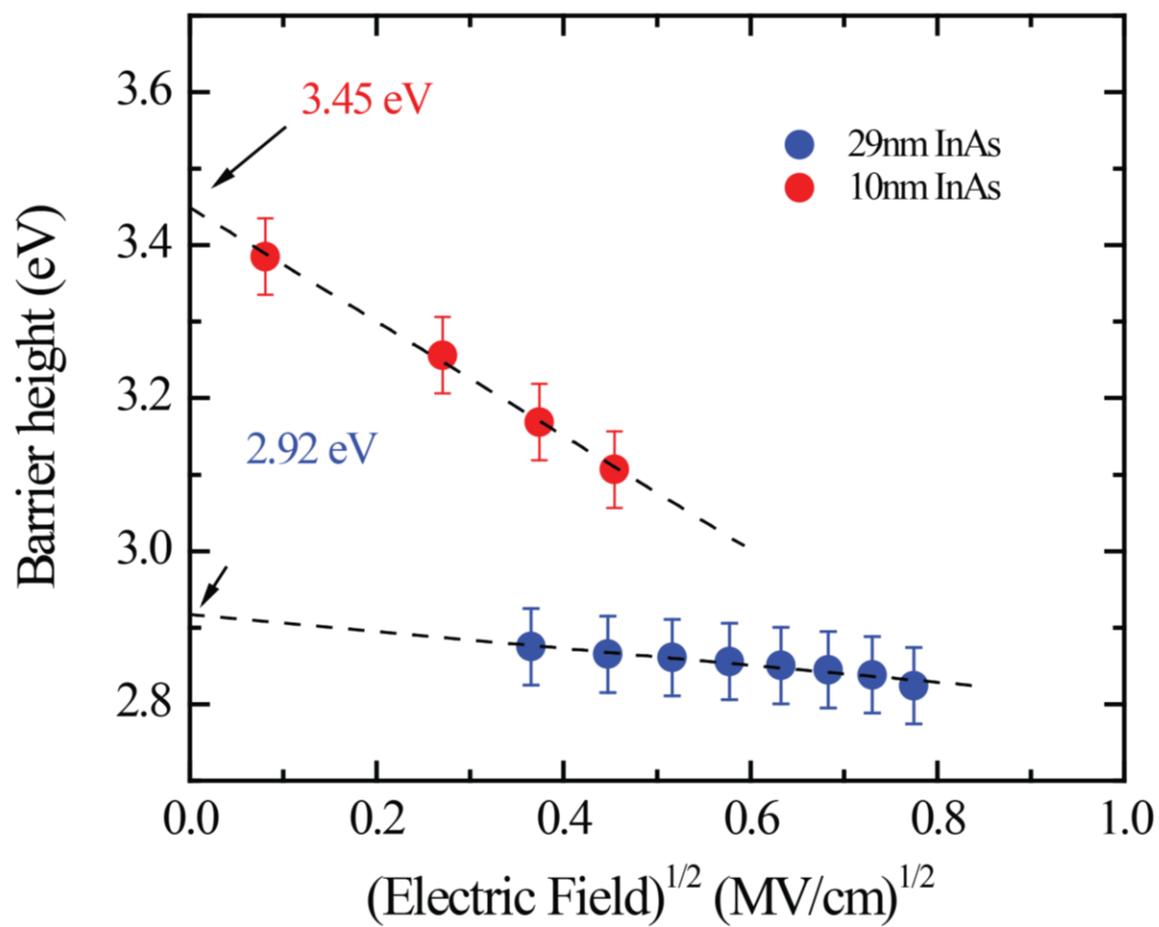

Figure 4

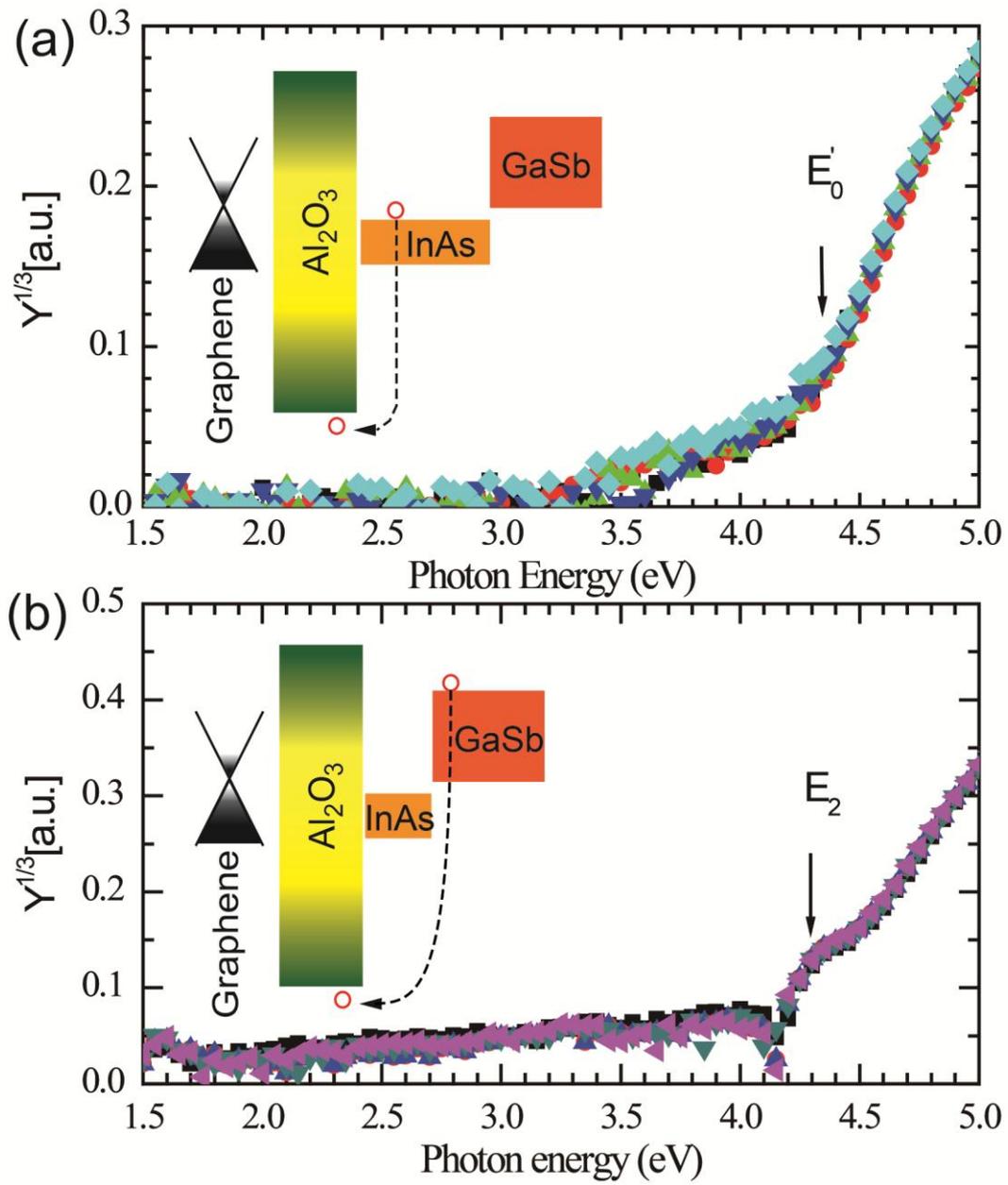

Figure 5

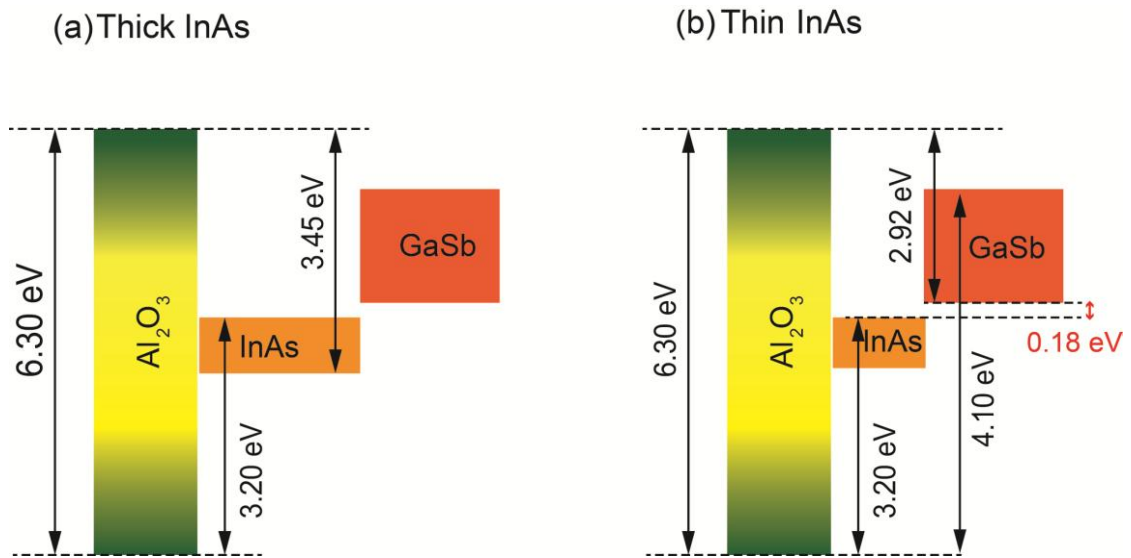

Figure 6